\documentclass[preprint,12pt]{elsarticle}

\usepackage{amssymb}

\journal{Chemical Physics Letters}

\begin{document}

\begin{frontmatter}

\title{Similarity law and critical properties in ionic systems.}
\author{Caroline Desgranges and Jerome Delhommelle\corref{cor}}
\cortext[cor]{Corresponding author. Email: jerome.delhommelle@und.edu}

\address{Department of Chemistry, University of North Dakota, Grand Forks ND 58202}

\begin{abstract}
Using molecular simulations, we determine the locus of ideal compressibility, or Zeno line, for a series of ionic compounds. We find that the shape of this thermodynamic contour follows a linear law, leading to the determination of the Boyle parameters. We also show that a similarity law, based on the Boyle parameters, yields accurate critical data when compared to the experiment. Furthermore, we show that the Boyle density scales linearly with the size-asymmetry, providing a direct route to establish a correspondence between the thermodynamic properties of different ionic compounds.
\end{abstract}

\begin{keyword}
Ionic fluids, Alkali Halides, Zeno line, Critical properties; Similarity Law. 
\end{keyword}

\end{frontmatter}
\section{Introduction}
Recent advances in the theory of vapor liquid equilibrium have led to the identification of new similarity laws, which allow for alternate pathways for the determination of the critical properties of fluids~\cite{apfelbaum2008new,kulinskii2010simple,kulinskii2014critical,apfelbaum2015wide}. Such theories rely on the prior determination of a contour known as the Zeno line, i.e. the line for which the compressibility factor $Z$ is equal to 1 ($Z=PV/RT=1$), that spans the subcritical and supercritical domain of the phase diagram~\cite{apfelbaum2013regarding}. Remarkably, in the case of model systems (i.e. for a Van der Waals fluid and the Lennard-Jones system)~\cite{nedostup2013asymptotic}, the Zeno line has been shown to be accurately modeled by a linear law that crosses the temperature and density axes at the Boyle temperature $T_B$ and density $\rho_B$, respectively. The Boyle parameters can then be used to obtain the critical properties through the following similarity law: 
\begin{equation}
{T_c \over T_B}+ {\rho_c \over \rho_B}=S
\label{Eq1}
\end{equation}
where $T_c$ and $\rho_c$ are the critical temperature and critical density, respectively, and $S$ is a similarity parameter, whose value depends on the type of intermolecular interactions taking place in the fluid~\cite{apfelbaum2009confirmation}.

This equation provides a direct route to determine the critical parameters when they are difficult to access  through experiments, such as e.g. for metals~\cite{apfelbaum2009predictions,apfelbaum2012estimate,Leanna} which exhibit very high critical temperatures. Recent work based either on experimental data or on molecular simulations have started to show the usefulness of Zeno line based approaches and its validity for a wide range of systems including metals and molecular fluids~\cite{kutney2000zeno,wei2013isomorphism,Landon,PartV,Parker}. Moreover, the Zeno line as well as the other thermodynamic regularities such as the line of ideal enthalpy, have emerged as a way to draw a correspondence between the thermodynamic states of supercritical fluids~\cite{brazhkin2011van,Abigail}. 

The determination of the critical properties of ionic systems has long proven to be especially challenging~\cite{yan2001phase,kim2005universality,Rane}. This is due to the strong Coulombic interactions that take place between ions and result in very high critical temperatures for molten salts, and also to the strong finite size effects in simulations of the critical region of the phase envelope. For such systems, it would be of particular interest to asses the validity of the similarity law of Eq.~\ref{Eq1}, and to test the reliability of approaches based on the linearity of the Zeno line to provide the critical parameters. Recent findings on charged Lennard-Jones systems have indeed indicated that the Zeno line is expected to depart from a straight line for strongly ionic systems, such as e.g. molten salts, in which the contribution of the long-range Coulombic interactions to the total interaction energy becomes very large~\cite{anashkin2017thermodynamic}. The aim of this work is to carry out an analysis of the behavior of the Zeno line for such systems, focusing here on the example of alkali halides, and to apply the results of this analysis to determine the critical parameters of ionic systems. Using molecular dynamics simulations over a wide range of state points, we determine the locus for the Zeno line for a series of alkali halides, and find that the Zeno line is accurately fitted by a linear law for subcritical temperatures (typically below 3000~K). This allows us to obtain the Boyle parameters for alkali halides and to show that a similarity law of the form given in Eq.~\ref{Eq1} can be reliably applied to determine the critical parameters for these systems. After determining the value of the constant $S$ of Eq.~\ref{Eq1} for alkali halides, we also show that the ratio $T_c:T_B$ and $\rho_c:\rho_B$, which are key in characterizing the asymmetry of the binodal, remain the same for the alkali halides studied in this work, thereby providing a way to evaluate the critical temperature and critical density. In addition, we quantify the impact of the size-asymmetry of the ions on the Boyle density and relate these findings to prior work focusing on the dependence of the critical density on size-asymmetry in restricted primitive models~\cite{yan2001phase,kim2005universality}. 

The paper is organized as follows. In the next section, we introduce the simulation models for the ionic systems as well as the simulation method used to determine the Zeno line, the Boyle parameters and the critical point. We then discuss the results obtained for the a series of alkali halides, starting with the determination of the Boyle parameters and the assessment of the similarity law for the well-studied system of $NaCl$. This allows us to find the value for the constant $S$ in the case of alkali halides, the relative values of the ratio of the critical to Boyle parameters and thus of the critical parameters. The results so obtained are tested against the simulation results on $KCl$ and against estimates for the critical temperature obtained from experimental data for a series of alkali halides. We then rationalize the relation between Boyle density and the size asymmetry between the ions and finally draw the main conclusion of this work in the last section.

\section{Simulation method}

We use the Born-Huggins-Mayer-Fumi-Tosi (BHMFT) potential to model alkali halides~\cite{tosi1964ionic}. The interaction between two ions $i$ and $j$ is given by
\begin{equation}
U(r_{ij})= {{z_i z_j e^2} \over {4 \pi D_0 r_{ij}}}+ A_{ij} \exp [B (\sigma_{i}+\sigma_{j} -r_{ij})] - {C_{ij} \over {r_{ij}^6}}- {D_{ij} \over {r_{ij}^8}}
\end{equation}
where $A_{ij}$, $B (\sigma_{i}+\sigma_{j})$, $C_{ij}$ and $D_{ij}$ are the parameters determined by Fumi and Tosi from the properties of solid alkali halides. Although it does not account for polarization effects, this potential has been shown to model correctly the structure, thermodynamics~\cite{guissani1994coexisting,rodrigues2007phase} and transport in molten salts~\cite{sindzingre1990computer,delhommelle2003shear,petravic2003conductivity}.

The Zeno line is the locus of the state points for which the compressibility factor ($Z=PV/RT$) is equal to 1. There are several ways to determine such state points~\cite{apfelbaum2013regarding,Abigail}. Here, we carry out a series of molecular dynamics (MD) simulations in the $(N,V,T)$ ensemble for each of the ionic compounds studied here (LiCl, NaCl, KCl, RbCl and CsCl) to capture the variation of the pressure as a function of the density of the system. From a practical standpoint, MD simulations are carried out on systems with a total of $N=512$~ions (256 cations and 256 anions) in a cubic cell of fixed volume $V$, with the periodic boundary conditions applied in the three directions. To keep the temperature constant, we use a Gaussian thermostat, which sets the time derivative of the kinetic temperature to zero and has been shown to yield equivalent results to Nose-Hoover thermostat~\cite{evans1985nose}. At a given temperature $T$, we keep $N$ constant and change $V$, so that we systematically vary the density by $0.05~g/cm^3$ increments. For each density $\rho$ (or, equivalently, each volume $V$), we determine the average value of the pressure $P$, through the virial expression. This yields the value of the $PV$ product for a given $T$, and thus the set $(\rho_z,T_z)$ for which the compressibility factor $Z=PV/RT$ is equal to $1$. We then repeat this process for temperatures ranging from $1500~K$ to $3000~K$ range, with a $250~K$ increment. Long-range interactions are calculated using Ewald sums with a cutoff set to $L/2$, in which $L$ is the boxlength, for the real space electrostatic interactions, a reciprocal wave vector cutoff set to $k_{max}=6 \times 2 \pi /L$ and a convergence acceleration factor $\kappa = 1.8 \pi / L$ (more details can be found in our prior work~\cite{delhommelle2003shear,petravic2003conductivity}).

\section{Results and Discussion}

\begin{figure}
\begin{center}
\includegraphics*[width=10cm]{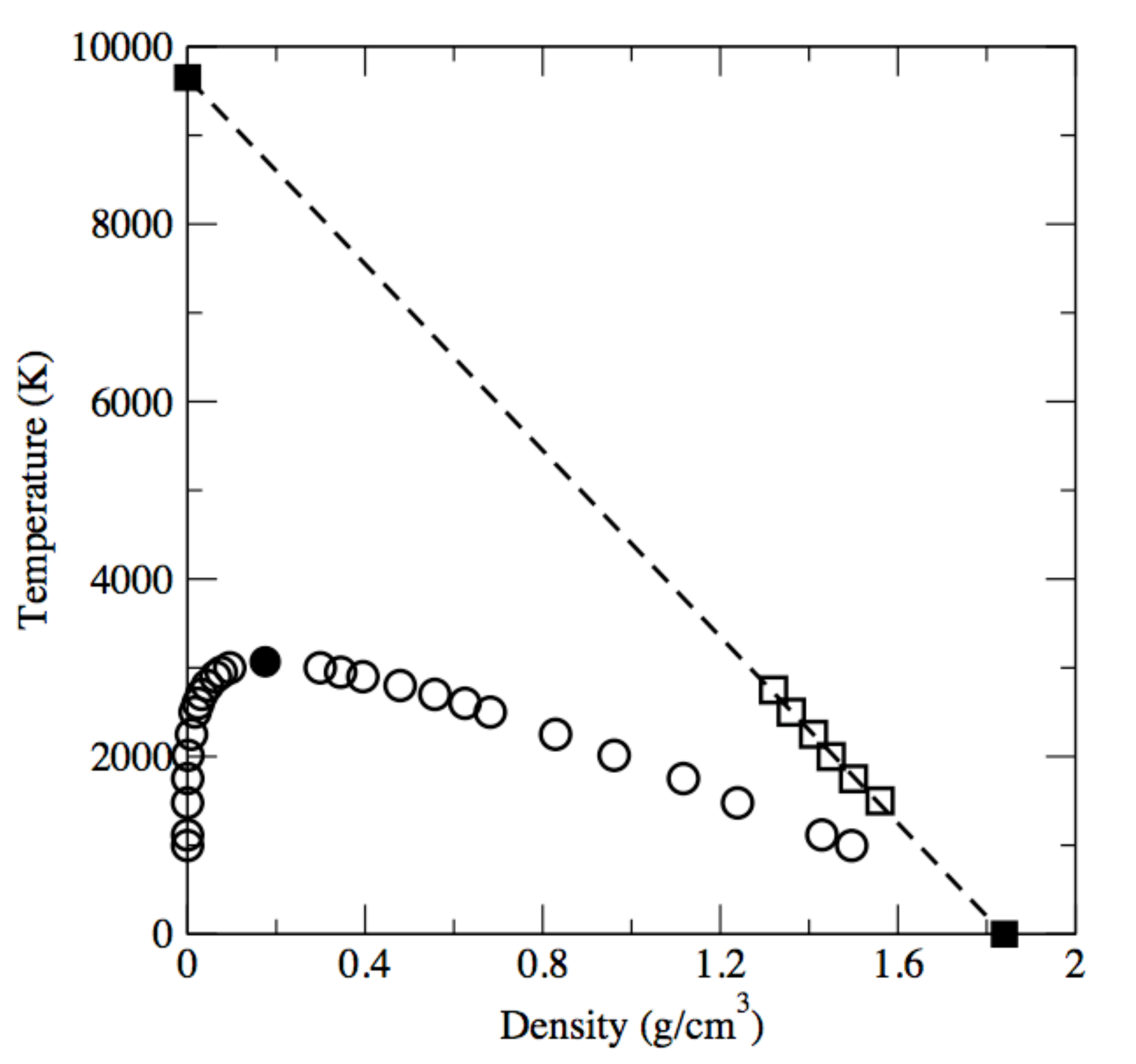}
\end{center}
\caption{Zeno line and binodal curve for NaCl. Simulation results for the Zeno line are shown as open squares (this work) while those for the binodal are shown as open circles (\cite{guissani1994coexisting}). Filled squares show the Boyle parameters and the filled circle corresponds to the critical point.}
\label{Fig1}
\end{figure}

We start by commenting on the results obtained for NaCl. The BHMFT model for NaCl has been extensively studied in simulations, and its structural, thermodynamic and transport properties have been shown to model accurately the experimental data. In particular, the critical density and temperature predicted by the model ($T_c=3068~K$ and $\rho_c=0.174~g/cm^3$)~\cite{guissani1994coexisting} have been shown to be in good agreement with the experimental data ($3000~K$ and $0.18~g/cm^3$)~\cite{kirshenbaum1962density}, and with the extrapolation of the experimental data proposed by Guissani and Guillot ($3300~K$ and $0.18~g/cm^3$). Since we use the BHMFT model for all alkali halides, together with the same system size as that used by Guissani and Guillot~\cite{guissani1994coexisting}, we take the critical data for the BHMFT model of NaCl as the starting point for our analysis. We therefore use this critical data as a reference to assess the analysis of the binodal based on the assumption of a straight Zeno line, as well as the applicability of the similarity law given in Eq.~\ref{Eq1}. We plot in Fig.~\ref{Fig1} the simulation data obtained along the Zeno line, together with the binodal curve previously obtained for this model by Guissani and Guillot~\cite{guissani1994coexisting}. The first question we set out to address is that of the linearity of the Zeno line for alkali halides. Using the simulations results, we find that the Zeno line is adequately fitted by a linear law showing that the Zeno line can be indeed taken to be a straight line for subcritical temperatures. This may seem {\it a priori} unexpected since the straightness of the Zeno line is known to break down for long-range interactions. We interpret this linear behavior as follows. For subcritical temperatures, the Zeno line is located in the liquid domain of the phase diagram, and generally involves high density liquids. This, in turn, implies that the long-range Coulombic interactions are greatly screened, given that the structure of the liquid can be described as successive layers of ions of opposite charge. This behavior is consistent with that previously observed by Anashkin {\it et al.} in their study of charged Lennard-Jones system~\cite{anashkin2017thermodynamic}, who observed large deviations of the Zeno line from a straight line at high, supercritical, temperatures and thus for highly dilute phases.

We fit the simulation results for NaCl with a linear law, as plotted in Fig.~\ref{Fig1}. This allows us to identify the Boyle parameters for NaCl, by taking the intercept of the linear law as the Boyle temperature $T_B=9650\pm150 K$, and the Boyle density as $\rho_B=1.84\pm0.04~g/cm^3$. From there, using the critical parameters for the BHMFT model used here for NaCl ($T_c=3068~K$ and $\rho_c=0.174~g/cm^3$), we can characterize the phase behavior of NaCl through the asymmetry of the binodal and determine the value taken by the constant $S$ in the similarity law of Apfelbaum and Vorob{'}ev. Specifically, we find that the ratio of the critical-to-Boyle temperatures is of $0.32\pm0.01$, and is much larger than the ratio of the critical-to-Boyle-density ($0.1\pm0.01$). This difference between the values taken by the 2 ratios indicates the strong asymmetry of the binodal curve exhibited by NaCl and, more generally, molten salts. Adding up these two ratios also provides the value of the constant used in the similarity law ($S=0.42\pm0.02$), leading to the following law for alkali halides:
\begin{equation}
{T_c \over T_B}+ {\rho_c \over \rho_B}=0.42
\label{Simsalt}
\end{equation} 

The value taken by the ratio $T_c$:$T_B$ is characteristic of the asymmetry of the binodal~\cite{Abigail}, and, like the similarity parameter $S$, is closely connected to the type of intermolecular interactions. We therefore assume that this ratio remains the same for all alkali halides. To assess this, we apply our analysis to the KCl system, modeled with the BHMFT potential, for which the binodal curve has been determined recently by Rodrigues and Fernandes~\cite{rodrigues2007phase}. We thus carry out a series of NVT simulations for temperatures ranging from $1500~K$ to $3000~K$ to identify the locus of the Zeno line and to determine the Boyle parameters. Then, using the estimate for $T_B$, we calculate the critical temperature $T_c$ using the $T_c:T_B$ ratio of 0.32 obtained for NaCl. Similarly, we calculate an estimate for the critical density $\rho_c$ from the Boyle density $\rho_B$ using the ratio of 0.1 for $\rho_c:\rho_B$ obtained for NaCl. We gather in Table~\ref{Tab1} the results obtained here for the Boyle parameters and for the critical parameters, together with the results obtained in previous work for the BHMFT model~\cite{rodrigues2007phase}, using a density expansion analysis. The critical parameters obtained in this work are in good agreement with those of prior work (within $20~K$ for the critical temperature and within $0.04~g/cm^3$ for the critical density), which confirms that the determination of the critical parameters from the Boyle parameters and the Zeno line can be reliably applied to determine the critical parameters of ionic compounds.

\begin{table}[hbpt]
\caption{Boyle and critical parameters for KCl ($T$ are given in $K$, and $\rho$ in $g/cm^3$).}
\begin{tabular}{|c|c|c|c|c|}
\hline
\hline
 &  $T_B$ &  $\rho_B$ & $T_c$ & $\rho_c$\\
\hline
\hline
Rodrigues~\cite{rodrigues2007phase} & - & -  & $2697 \pm 16$ & $0.216 \pm 0.009$ \\
This work & $8490 \pm 80$ & $1.77 \pm 0.04$ & $2717 \pm 26$ & $0.177 \pm 0.004$ \\
\hline
\hline
\end{tabular}
\label{Tab1}
\end{table}

\begin{figure}
\begin{center}
\includegraphics*[width=14cm]{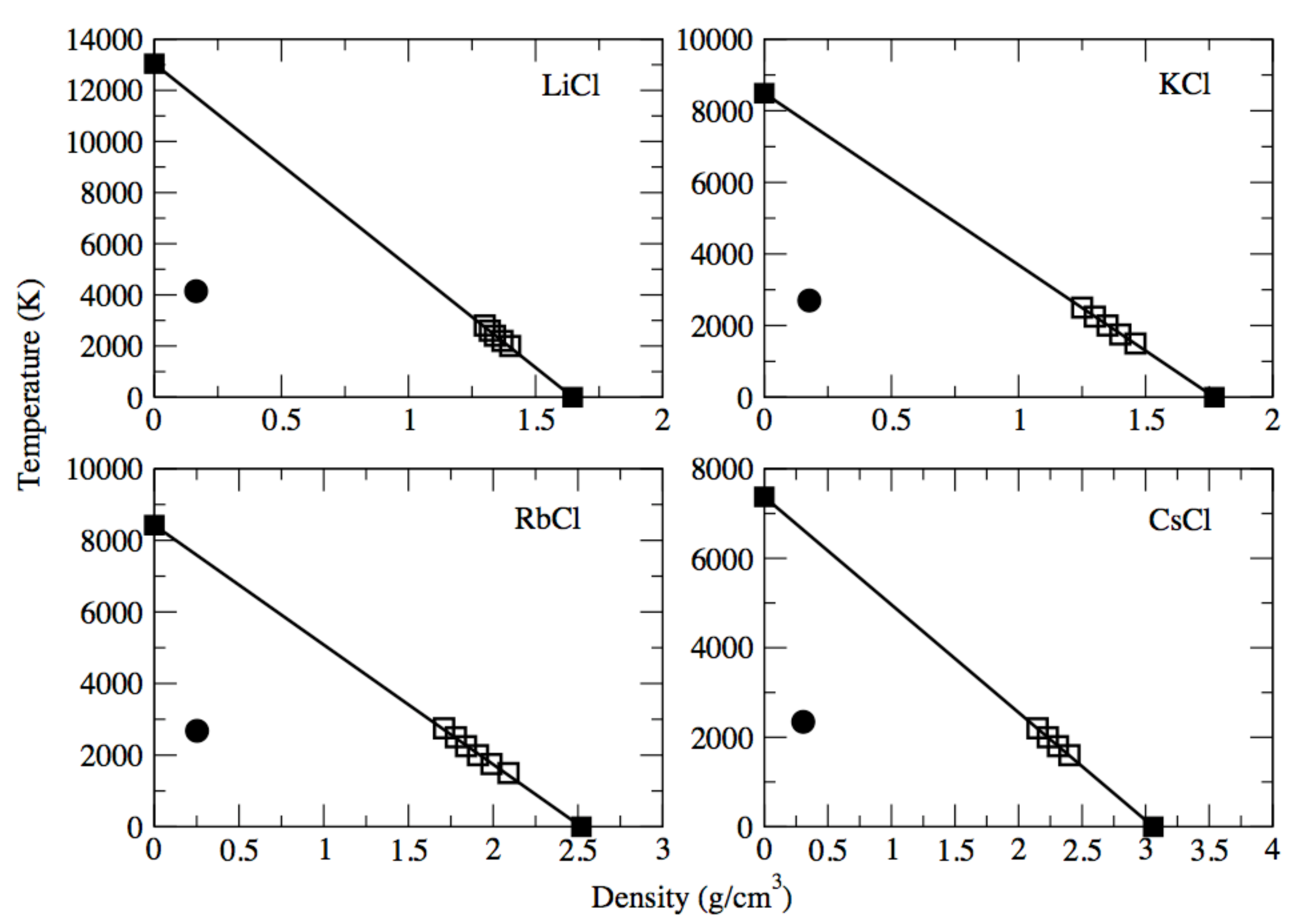}
\end{center}
\caption{Zeno line and estimates for the Boyle and critical parameters for LiCl, KCl, RbCl and CsCl (same legend as Fig.~\ref{Fig1}).}
\label{Fig2}
\end{figure}

We now move on to the other alkali halides considered in this work. To gain insight into the impact of size-asymmetry on the Boyle parameters, we focus on ionic compounds involving chloride anions and examine the whole range of sizes for alkali cations from Lithium to Cesium. Through successive MD simulations, we determine the locus for the Zeno line for LiCl, KCl, RbCl and CsCl. We show the resulting plots in Fig.~\ref{Fig2}, together with the critical points using the similarity law of Eq.~\ref{Simsalt}, with $T_c:T_B$ and $\rho_c:\rho_B$ of 0.32 and 0.1, respectively. To provide a comparison with experimental data on these systems, we compile the experimental data on the surface tension of ionic compounds obtained in prior work~\cite{sato1990surface} and fit the data with Guggenheim's law to determine the critical temperature~\cite{guggenheim1945principle}
\begin{equation}
\gamma = \gamma_0 (1- {T \over T_c})^{11/9}
\label{Gug}
\end{equation} 
where $\gamma$ is the surface tension and $\gamma_0$ and $T_c$ are fitting parameters. This approach has the advantage of providing a consistent set of data for the critical temperatures, that will effectively capture the dependence of $T_c$ on the type of ionic compound considered. We add that this approach has been shown to be a reliable way to obtain estimates for the critical temperature in prior work on ionic liquids~\cite{rebelo2005critical}.

We summarize the results from this work for the Boyle and critical parameters, together with the critical temperature estimated from the experimental data, in Table~\ref{Tab2}. The approach used here, which relies on the prior determination of the thermodynamic locus for the Zeno line, yields critical temperatures that are overall in good agreement with the experimental data. We find deviations within $3$~\% for all compounds from NaCl to CsCl, with the notable exception of LiCl for which the deviation rises to $16$~\%. Most interestingly, the method used in this work is able to capture the monotonic decrease of $T_c$ with the increase in size for the cation, thereby showing that this approach is sensitive enough to the details of the ionic interaction potentials to model accurately this trend.

\begin{table}[hbpt]
\caption{Comparison of Boyle and critical parameters for alkali halides (this work) to the critical parameters estimated from the experimental data on the surface tension~\cite{sato1990surface}  ($T$ are given in $K$, and $\rho$ in $g/cm^3$).}
\begin{tabular}{|c|c|c|c|c|c|}
\hline
\hline
  &  $T_B$ &  $\rho_B$ & $T_c$ & $\rho_c$ & $T_c$~\cite{sato1990surface} \\
\hline
\hline
LiCl & $13043 \pm 250$ & $1.645 \pm 0.02$  & $4173 \pm 80$ & $0.165 \pm 0.002$ & $3454$ \\
NaCl & $9650 \pm 150$ & $1.84 \pm 0.04$ & $3088 \pm 50$ & $0.184 \pm 0.004$ & $3044$ \\
KCl & $8490 \pm 80$ & $1.77 \pm 0.04$ & $2717 \pm 26$ & $0.177 \pm 0.004$ & $2691$ \\
RbCl & $8424 \pm 80$ & $2.52 \pm 0.03$ & $2695 \pm 28$ & $0.252 \pm 0.003$ & $2525$ \\
CsCl & $7371 \pm 70$ & $3.06 \pm 0.04$ & $2359 \pm 26$ & $0.306 \pm 0.004$ & $2464$ \\
\hline
\hline
\end{tabular}
\label{Tab2}
\end{table}

Prior work on the effect of size-asymmetry on the critical density in restricted primitive models~\cite{yan2001phase,kim2005universality} has highlighted systematic variations of $\rho_c$ as a function of size-asymmetry. We thus focus on the variations of the Boyle density as a function of size-asymmetry, across the range of alkali halides studied here. We start by defining the size-asymmetry parameter $\lambda = \sigma_+ / \sigma_-$, where $\sigma_+$ and $\sigma_-$ are the atomic radii of the cation and anion (here, the chloride anion), respectively, as given in BHMFT potential~\cite{tosi1964ionic}. To compare the results obtained for the various alkali chlorides, we define a reduced Boyle density and use the following scaling
\begin{equation}
n^*_B= { \rho_B {\sigma_-}^3 \over (m_+ + m_-)}
\label{reducedrb}
\end{equation} 
where $\rho_B$ is the Boyle density (in $g/cm^3$), while $m_+$ and $m_-$ are the masses for the cation and the chloride anions, respectively.

\begin{figure}
\begin{center}
\includegraphics*[width=10cm]{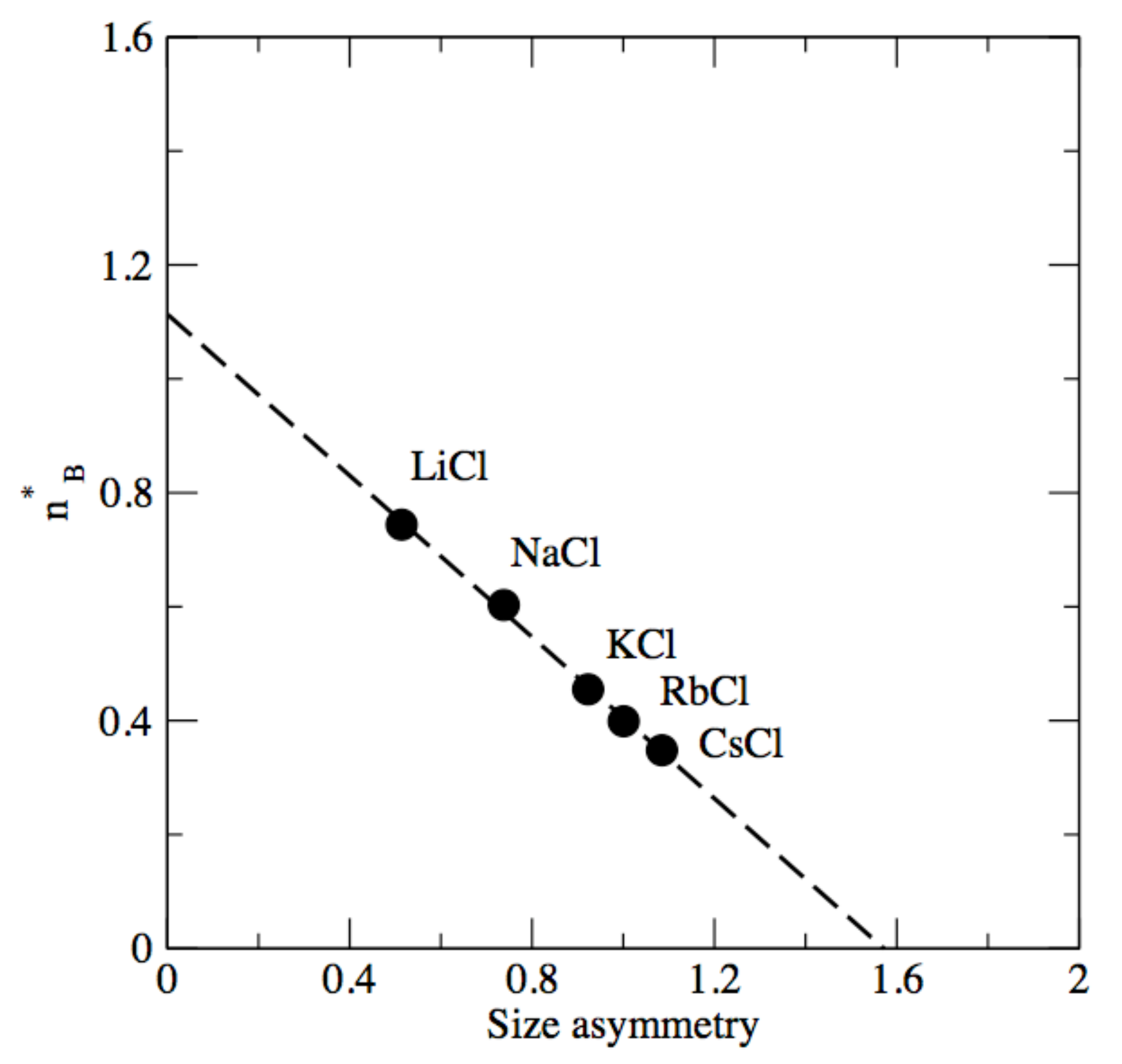}
\end{center}
\caption{Variation of the reduced Boyle density as a function of size asymmetry. Boyle densities estimated from the simulations are shown are filled squares for the alkali chloride series, and the dashed line shows a linear variation as a function of the size asymmetry.}
\label{Fig3}
\end{figure}

We show in Fig.~\ref{Fig3} the resulting reduced Boyle densities $n^*_B$ against the size-asymmetry parameter. The reduced Boyle density is found to decrease with $\lambda$, as a result of the decrease in the size of the cation while the anion remains the same. Fig.~\ref{Fig3} shows that the simulation results for the reduced Boyle densities align remarkably well for the alkali chloride series, and can be very accurately fitted by a linear law. We obtain the following fit: $n^*_B (\lambda) = (1.114 \pm 0.009) - (0.71 \pm 0.02) \times \lambda$. This shows that the mapping of the thermodynamic properties, most notably of the binodal through the law of corresponding states, can be extended to other regions of the thermodynamic space, as shown here in the case of the Zeno line of different ionic compounds through the scaling proposed here in Eq.~\ref{reducedrb}. 
 
\section{Conclusions}
In this work, we determine the Zeno line, together with the associated Boyle density and temperature, for a series of ionic compounds, and assess the ability of similarity laws, based on the Boyle parameters, to predict accurately the critical temperature and critical density for ionic systems. For this purpose, we carry out series of molecular dynamics simulations of alkali chloride compounds and determine the locus for the Zeno line, where the compressibility factor is equal to 1. Our main conclusions are as follows. The Zeno line is well fitted by a linear law for temperatures below 3000~K, which allows us to determine the Boyle parameters for these compounds. Using NaCl as a reference compound, we are able to: (i) identify a similarity law, based on the Boyle parameters, leads to accurate critical parameters, (ii) determine the value of the constant in the similarity law for alkali halides, and (iii) quantify the asymmetry of the binodal through the ratios of the critical parameters to the Boyle parameters. Testing this approach on a series of alkali chloride compounds yields accurate critical temperatures, when compared to the experimental data, and displays the correct trend as a function of the nature of the cation involved. Furthermore, we find that the reduced Boyle density exhibits a linear variation as a function of the size-asymmetry parameter, thereby providing evidence that a mapping of the thermodynamic properties of different ionic compounds can be established through this scaling.\\

{\bf Acknowledgements}
Partial funding for this research was provided by NSF through CAREER award DMR-1052808.

\bibliographystyle{model1a-num-names}

\bibliography{ZenoNaCl}

\end{document}